\documentclass{aa}  

\usepackage{graphicx}
\usepackage{txfonts}
\usepackage{grffile}
\usepackage{longtable}
\def\nh3{NH$_{3}$}
\def\kms{km~s$^{-1}$}

\def\Vlsr{$V_{\rm LSR}$}
\def\Jyb{Jy~beam$^{-1}$}

\def\G24{G24.78$+$0.08}
\def\HII{H{\sc ii}}

\newcommand{\ms}{$M_{\odot}$}
\newcommand{\ls}{$L_{\odot}$}

\newcommand{\degree}{$^{\circ}$}

\usepackage{color}

\begin{document} 

 %  \title{10--10$^3$~AU View of Outflows from high-mass YSOs.}
   \title{Protostellar Outflows at the EarliesT Stages (POETS). IV. Statistical properties of the 22~GHz H$_2$O masers}
    
  \subtitle{}

  \titlerunning{Statistical properties of H$_2$O masers.}
   
   \author{L. Moscadelli\inst{1}
          \and
          A. Sanna\inst{2,3}
          \and
          C. Goddi\inst{4,5}
          \and
          V. Krishnan\inst{1,6}
          \and
          F. Massi\inst{1}
          \and
          F. Bacciotti\inst{1}}
  
   \institute{INAF-Osservatorio Astrofisico di Arcetri, Largo E. Fermi 5, 50125 Firenze, Italy \\
              \email{mosca@arcetri.astro.it}
             \and 
             INAF-Istituto di Radioastronomia \& Italian ALMA Regional Centre, Via P. Gobetti 101, I-40129 Bologna, Italy
             \and
              Max-Planck-Institut f\"{u}r Radioastronomie, Auf dem H\"{u}gel 69, 53121 Bonn, Germany 
             \and
              Leiden Observatory, Leiden University, PO Box 9513, 2300 RA Leiden, The Netherlands 
             \and
             Department of Astrophysics/IMAPP, Radboud University, PO Box 9010, 6500 GL, Nijmegen, The Netherlands  
             \and
             South African Radio Astronomy Observatory (SARAO), 2 Fir street, Black River Park, Observatory, Cape Town, 7925, South Africa            
             }

   \date{}

% \abstract{}{}{}{}{} 
% 5 {} token are mandatory
 
  \abstract
  % leave it empty if necessary  
  {22~GHz water masers are the most intense and widespread masers in star-forming regions. They are commonly associated with protostellar winds and jets emerging from low-~and~high-mass young stellar objects (YSO).}
   {We wish to perform for the first time a statistical study of the location and motion of individual water maser cloudlets, characterized by typical sizes that are within a few~au, with respect to the weak radio thermal emission from YSOs.}
   {For this purpose, we have been carrying out the  Protostellar Outflows at the EarliesT Stages (POETS) survey of a sample (38) of high-mass YSOs. The 22~GHz water maser positions and three-dimensional (3D) velocities were determined through multi-epoch Very Long Baseline Array (VLBA) observations with accuracies of a few milliarcsec (mas) and a few \kms, respectively. The position of the ionized core of the protostellar wind, marking the YSO, was determined through sensitive radio continuum, multi-frequency Jansky Very Large Array (JVLA) observations with a typical error of \ $\approx$20~mas.  }
   {The statistic of the separation of the water masers from the radio continuum shows that \ 84\% \  of the masers are found within 1000~au from the YSO and \ 45\% \ of them are within \ 200~au. Therefore, we can conclude that the 22~GHz water masers are a reliable proxy for locating the position of the YSO.
The distribution of maser luminosity is strongly peaked towards low values, 
indicating that about half of the maser population is still undetected with the current Very Long Baseline Interferometry (VLBI) detection thresholds of  \ 50--100~m\Jyb. Next-generation, sensitive (at the nJy level) radio interferometers will have the capability to exploit these weak masers for an improved sampling of the velocity and magnetic fields around the YSOs.
The average direction of the water maser proper motions provides a statistically-significant estimate for the orientation of the jet emitted by the YSO: 55\% \ of the maser proper motions are directed on the sky within an angle of \ 30\degree\ from the jet axis. Finally, we  show that our measurements of 3D maser
velocities statistically support models in which water maser emission arises from planar shocks with propagation direction close to the plane of the sky.}
   {}


   \maketitle
%
%-------------------------------------------------------------------

% Acronyms:  PM 3D mas DW VLBI rms YSO

\section{Introduction}
\label{intro}

% Connection between water masers and star-formation. Comparison with previous water maser surveys in star-forming regions.
The 22~GHz water maser 6$_{1,6}$-5$_{2,3}$ rotational transition is the most intense emission line at radio frequency and the most widespread interstellar maser. In star-forming regions, the average isotropic luminosity of detected water masers is \ $\sim$10$^{-5}$~\ls\ \citep[][see Table~5]{Ang96b} and in extreme cases, such as in the strong source W49N, it can be as high as  \ $\sim$1~\ls \ \citep{Eli89}. From the time of its discovery by \citet{Che69}, the water maser emission has been surveyed using both single-dish \citep{Fur01,Fel07,Urq11,Wal11} and, more recently, connected interferometers \citep{Beu02a,Wal14,Tit16,Kim19}, with angular resolutions ranging from \ $\sim$1\arcsec \ to \ $\sim$1\arcmin. These surveys have derived the global properties of the emission (luminosity, extent in LSR velocity (\Vlsr) and the time variability) of the interstellar water masers, in addition to studying the correlation with independent star-formation markers, particularly with other maser types. Very long baseline interferometry (VLBI) observations of specific sources, achieving milliarcsecond (mas) angular resolution, allow us to resolve the emission and determine the spatial and spectral characteristics of individual maser cloudlets\ and measure their three-dimensional (3D) motion \citep{God06b}. The VLBI results in most cases indicate a physical association of the water masers with protostellar winds and jets emerging from low-~and~high-mass young stellar objects (YSO) \citep{Tor03,Mos05,God06a,San12,Bur16,Hun18}.
% BeSSeL and the chance of a VLBI survey of water masers. 

Models of the 22~GHz water masers predict that the emission originates in warm and dense, shocked gas. Two different classes of shocks are considered: slow ($\le$40~\kms) non-dissociative C-type \citep{Kau96}, and fast ($\ge$40~\kms) dissociative J-type \citep{Eli92,Hol13} shocks. The post-shock densities for water maser operation, H$_2$ number density (n$_{\rm{H}_2}$) $\sim$$10^8$--$10^9$~cm$^{-3}$, are similar; however, in C~shocks molecules are not dissociated and the masing gas attains high temperatures of \ 1000--2000~K, whereas in J-shocks, the H$_2$O molecules reform in a post-shock temperature plateau of \ 400--500~K. Shocks are a natural by-product of winds and jets, caused either by a velocity or density change internal to the flow, or
by interaction of the flow with the surrounding medium. The link between the water masers and the outflows from YSOs provided by the observations is, therefore, consistent with the shock models, although a more quantitative test of the models, in particular discerning between C-~and~J-shocks, is often hampered by insufficient knowledge about the flow geometry and its speed.

We have been carrying out the  Protostellar Outflows at the EarliesT Stages (POETS) survey with the goal of imaging the inner outflow scales of \ 10-100~au \  in a statistically-significant sample (38\footnote{This number includes the two YSOs \ IRAS~20126$+$4104 \ and \ AFGL~5142 \ studied via multi-epoch VLBI observations by our group in the past, which have been added to the POETS sample of 36 targets (see Table~\ref{wat_kin}).}) of luminous YSOs, targeting both the molecular and ionized components of the outflows. The outflow kinematics is studied at mas scales through Very Long Baseline Array (VLBA) observations of the 
22~GHz water masers belonging to the Bar and Spiral Structure Legacy survey (BeSSeL; see below). We have employed the Jansky Very Large Array (JVLA) at C- (6~GHz), Ku- (13~GHz), and K-band (22~GHz) in the A-~and~B-Array configurations (FWHM beams of \ 0\farcs1--0\farcs4) to determine the spatial structure and the spectral index of the radio continuum emission and to address its nature. 

The BeSSeL survey is a key project of the VLBA, whose main goal is to derive the structure and kinematics of the Milky Way by measuring accurate positions, distances (via trigonometric parallaxes) and proper motions (PM) of 6.7~GHz methanol and 22~GHz water masers in hundreds of high-mass star forming regions distributed over the Galactic Disk \citep{Rei14,Rei19}. The BeSSeL observations provide absolute positions and velocities of individual maser cloudlets with accuracies of a few mas and a few \kms, respectively, for hundreds of YSOs. They have an enormous value for star-formation studies since they afford for the first time a VLBI survey of the properties of the most widespread interstellar masers based on a large, homogeneous dataset.

\citet{Mos16} (hereafter, Mos16) presented the first results of the POETS survey for a pilot sample of 11~targets, also describing the target selection, observations, and data analysis.  \citet{San18} (hereafter Paper~I) reported on and interpreted the radio continuum data of the whole sample, while  \citet{San19b} (hereafter Paper~II) examined the radio synchrotron jet associated with the high-mass YSO \ G035.02$+$0.35.
\citet{Mos19b} (hereafter Paper~III) completed the combined analysis of the radio continuum and water maser observations for all the targets, with a particular focus on the water maser kinematics.
The main result of Paper~III is that the 3D velocity distribution of the water masers near the YSO in all the sources of the sample can be interpreted in terms of a single physical scenario:
% a jet emerging from a disk-wind (DW). 
a disk-wind (DW). The observed masers are produced in the different regions of the DW, from the axial collimated jet portion 
to the wide-angle outer layers. 

In this paper, we report on the statistics of the water maser properties and examine global correlations in the observed physical quantities.
The distribution of water maser positions, luminosities, and 3D-velocity orientations and amplitudes is presented in Sect.~\ref{obs_res}. 
In Sect.~\ref{discu}, we discuss the use of the H$_2$O masers as a tool for star-formation studies. 
Our conclusions are presented in Sect.~\ref{conclu}.

%--------------------------------------------------------------------
\begin{figure*}
\centering
\includegraphics[width=\textwidth]{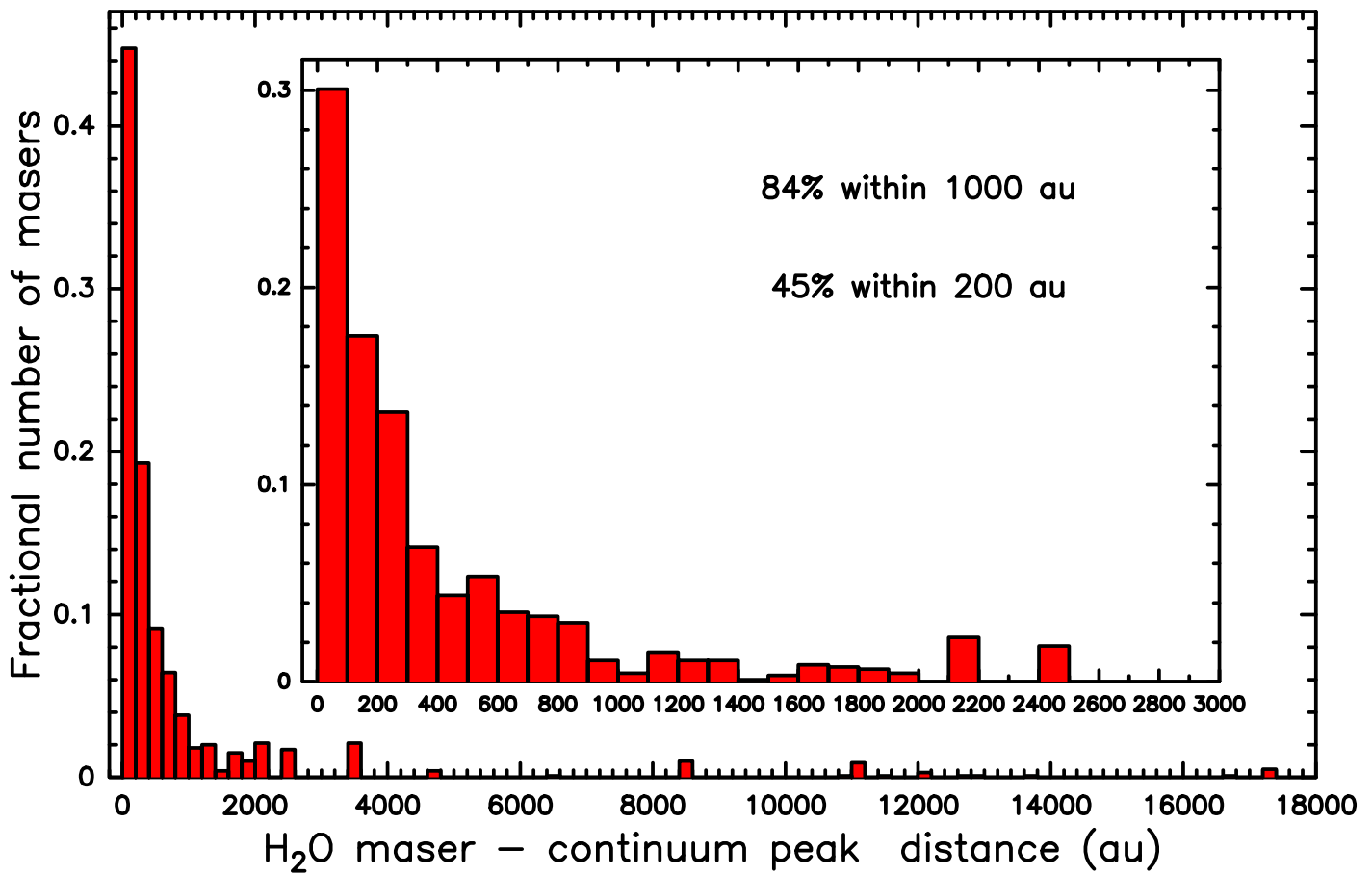}
\caption{Histograms of the sky-projected distance of water masers from the radio continuum peak, for all the observed masers in the POETS targets ({\it main}) and for a maximum continuum-maser separation of 3000~au ({\em inset}). Considering all the observed masers, the target-average distance error and corresponding standard deviation are 182~and~16~au, respectively. Selecting masers within 3000~au from the radio continuum peak, the target-average distance error and corresponding standard deviation are  \ 108~and~17~au, respectively. Correspondingly, we used bins of \ 200 and 100~au for the histograms in the main plot and in the inset, respectively. The histogram values are normalized by the total number of selected water masers. }
\label{histo_rad}
\end{figure*}

\begin{figure*}
\centering
\includegraphics[width=1.0\textwidth]{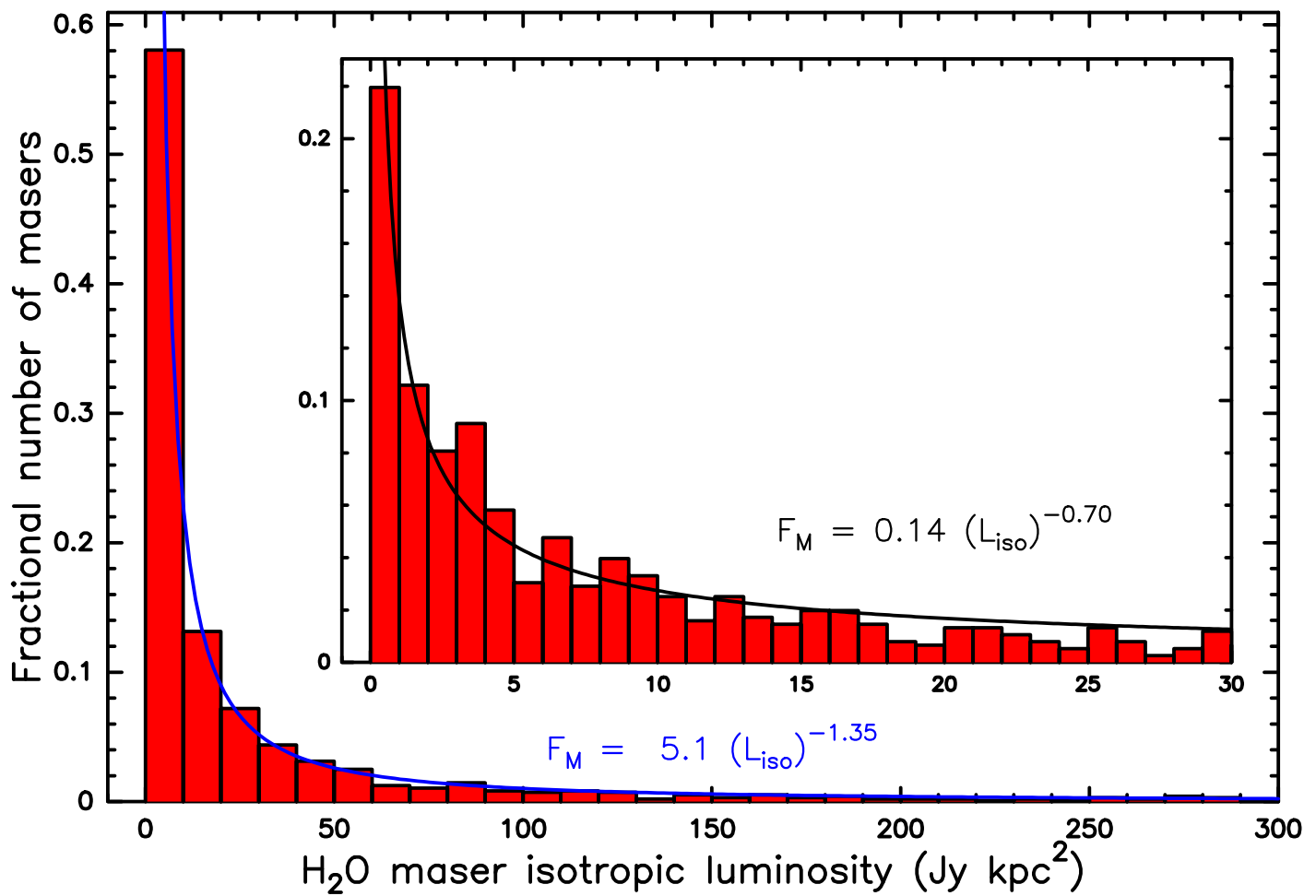}
\caption{Histograms of the maser isotropic luminosity for a maximum maser luminosity of \ 300~Jy~kpc$^2$ ({\it main}) \ and \ 30~Jy~kpc$^2$ ({\em inset}). Less than 4\% of the observed masers have an isotropic luminosity higher than \ 300~Jy~kpc$^2$. The bins are 10 and 1~Jy~kpc$^2$ for the histograms in the main plot and in the inset, respectively. The histogram values are normalized by the total number of selected water masers. The continuous blue and black  lines show the power-law fits of the values of the histograms in the main plot and in the inset, respectively, and the best-fit expressions are also reported using the same colors of the corresponding fits.}
\label{histo_lum}
\end{figure*}

%-------------------------------------------------------------
%

%--------------------------------------------------------------------

%\begin{table*}
\longtab{
%\begin{landscape}
\begin{longtable}{l c c c c r c}       
\caption{\label{wat_kin} Water masers and radio continuum}\\             % title of Table
%\label{wat_kin}      % is used to refer this table in the text
%\centering                          % used for centering table
%\begin{tabular}{c c c c c c c c c c c c c c c c c c c c}        % centered columns (4 columns)
\hline \hline                 % inserts double horizontal lines
%\noalign{\smallskip}
Source &     d     &  L$_{\rm bol}$         &  \multicolumn{2}{c}{YSO continuum\tablefootmark{a}}   &     N$_{\rm mas}$  & \multicolumn{1}{c}{Maximum radius} \\  % table heading
       &           &                        &            RA~(J2000)     &           DEC~(J2000)    &      &   \\
       &    (kpc)  &  (L$_{\odot}$)         &       ($\mathrm{h}$ \; \ $\mathrm{m}$ \; \ $\mathrm{s}$)     &  (\degr \; \; $^{\prime}$ \; \; $^{\prime\prime}$)  &    &  (au) \\
%\noalign{\smallskip}
\hline 
\endfirsthead
\caption{continued.}\\
\hline \hline                 % inserts double horizontal lines
%\noalign{\smallskip}
Source &     d     &  L$_{\rm bol}$         &  \multicolumn{2}{c}{YSO continuum\tablefootmark{a}}   &     N$_{\rm mas}$  & \multicolumn{1}{c}{Maximum radius} \\  % table heading
       &           &                        &            RA~(J2000)     &           DEC~(J2000)    &      &   \\
       &    (kpc)  &  (L$_{\odot}$)         &       ($\mathrm{h}$ \; \ $\mathrm{m}$ \; \ $\mathrm{s}$)     &  (\degr \; \; $^{\prime}$ \; \; $^{\prime\prime}$)  &    &  (au) \\
%\noalign{\smallskip}
\hline 
\endhead
\hline
\endfoot
G005.88$-$0.39 &   2.99$\pm$0.18 &   5.8 $\times$ 10$^4$  &  18 00 30.440 &  $-$24 04 00.90   &  26 &  16708   \\
G009.99$-$0.03\tablefootmark{b}  &   {\it 5.0 }    &   {\it 1.2 $\times$ 10$^4$}       &  18 07 50.117 &  $-$20 18 56.49  &  16 &   801    \\
G011.92$-$0.61 &   3.37$\pm$0.35 &   1.2 $\times$ 10$^4$  &  18 13 58.112 & $-$18 54 20.19  & 27 &   776  \\
G012.43$-$1.12\tablefootmark{b}  &   {\it 3.7 }    &   {\it 4.2 $\times$ 10$^4$}    &  18 16 52.161 &  $-$18 41 43.94  &    3 &  227   \\
G012.68$-$0.18 &   2.40$\pm$0.18 &   5.7 $\times$ 10$^3$  & 18 13 54.750 & $-$18 01 46.58    &       21 &  630   \\
G012.90$-$0.24  &   2.45$\pm$0.15 &   8.6 $\times$ 10$^2$   &  18 14 34.428 &   $-$17 51 51.80  &    2 &  130  \\
G014.64$-$0.58 &   1.83$\pm$0.07 &   1.1 $\times$ 10$^3$   &  18 19 15.546  & $-$16 29 45.83 &   5 &  185   \\
G016.58$-$0.05 &   3.58$\pm$0.30 &   1.3 $\times$ 10$^4$    &  18 21 09.125 & $-$14 31 48.65  &    35 &  2179    \\
G026.42$+$1.69\tablefootmark{b} &   {\it 3.1 } &   {\it 9.0 $\times$ 10$^3$}    &  18 33 30.508 & $-$05 01 01.95  &    7 &  275  \\
G031.58$+$0.08 &   4.90$\pm$0.72 &   2.0 $\times$ 10$^4$    &  18 48 41.613 &   $-$01 09 57.70  &   11 &  11484     \\
G035.02$+$0.35 &   2.33$\pm$0.22 &   1.0 $\times$ 10$^4$ & 18 54 00.649  & $+$02 01 19.41  &     27  &   621 \\
G049.19$-$0.34 &   5.29$\pm$0.20 &   6.0 $\times$ 10$^3$    &   19 22 57.771  & $+$14 16 09.91  &  11 &  254    \\
G074.04$-$1.71 &   1.59$\pm$0.05 &   {\it 3.7 $\times$ 10$^2$}   &  20 25 07.117 & $+$34 49 57.53   &  29 &   586   \\
G075.76$+$0.34 &   3.51$\pm$0.28 &   1.4 $\times$ 10$^4$   &  20 21 41.093 & $+$37 25 29.19  &   19 &  453  \\
G075.78$+$0.34 &   3.72$\pm$0.43 &   1.1 $\times$ 10$^4$   &  20 21 44.013 & $+$37 26 37.51  &    61 &  2439  \\
G076.38$-$0.62 &   1.30$\pm$0.09 &   {\it 1.4 $\times$ 10$^4$ }   &   20 27 25.477 & $+$37 22 48.42 &  50 &  115   \\
G079.88$+$1.18 &   1.61$\pm$0.07 &   8.6 $\times$ 10$^2$    &   20 30 29.145  &  $+$41 15 53.54   &  6 &  178  \\
G090.21$+$2.32 &   0.67$\pm$0.02 &   2.7 $\times$ 10$^1$   &  21 02 22.703  &  $+$50 03 08.27    &  15 &  38  \\
G092.69$+$3.08  &   1.63$\pm$0.05 &   ({\it 4.7 $\times$ 10$^3$})    &   21 09 21.714  &  $+$52 22 37.02  &  50 & 275    \\
G097.53$+$3.18--M &   7.52$\pm$0.96 &   ({\it 8.8 $\times$ 10$^4$}) &  21 32 12.451 & $+$55 53 49.76   &    82 &   1923      \\
G100.38$-$3.58 &   3.44$\pm$0.10 &   {\it 8.5 $\times$ 10$^3$}   &  22 16 10.365 & $+$52 21 34.14   &  49 &  755       \\
G105.42$+$9.88\tablefootmark{c}~3A &   0.89$\pm$0.05 &  ({\it 5.8 $\times$ 10$^2$})   &   21 43 06.474  &  $+$66 06 54.98  &   2 &  15    \\
G105.42$+$9.88\tablefootmark{c}~3B &   0.89$\pm$0.05 &   ({\it 5.8 $\times$ 10$^2$})  &   21 43 06.460  &  $+$66 06 55.18  &  72 & 23 \\
G108.20$+$0.59 &   4.37$\pm$0.53 &   2.1 $\times$ 10$^3$   & 22 49 31.465 &  $+$59 55 41.86  &       39 &  452   \\
G111.24$-$1.24 &   3.47$\pm$0.53 &   {\it 1.0 $\times$ 10$^4$}   &   23 17 20.891 &  $+$59 28 47.60  &   21 &  3575           \\
G111.25$-$0.77 &   3.40$\pm$0.18 &   {\it 5.0 $\times$ 10$^3$}  & 23 16 10.331 & $+$59 55 28.63    &    57 &  1268   \\
G160.14$+$3.16 &   4.10$\pm$0.10 &   {\it 8.4 $\times$ 10$^3$}   &  05 01 39.914  &  $+$47 07 21.58 &   5 &  17304  \\
G168.06$+$0.82 &   7.69$\pm$2.37 &   1.6 $\times$ 10$^4$    &  05 17 13.743  & $+$39 22 19.85  &    15 &   1849      \\
G176.52$+$0.20 &   0.96$\pm$0.02 &   1.5 $\times$ 10$^2$   &   05 37 52.136  & $+$32 00 03.93  &  71 &  177  \\
G182.68$-$3.27 &   6.71$\pm$0.50 &  {\it 8.6 $\times$ 10$^2$}        &   05 39 28.419 &  $+$24 56 32.16  & 3 & 347    \\
G183.72$-$3.66 &   1.75$\pm$0.04 &   {\it 9.7 $\times$ 10$^2$}   &   05 40 24.230 & $+$23 50 54.68   &   22 & 117 \\
G229.57$+$0.15\tablefootmark{d}~1 &   4.52$\pm$0.29 &   2.2 $\times$ 10$^3$   &  07 23 01.845 &  $-$14 41 32.79   &   2 & 161     \\
G229.57$+$0.15\tablefootmark{d}~2 &   4.52$\pm$0.29 &   2.2 $\times$ 10$^3$   &  07 23 01.804 &  $-$14 41 32.94   &   26 &  389    \\
G236.82$+$1.98 &   3.36$\pm$0.20 &  {\it 2.3 $\times$ 10$^3$ }   &  07 44 28.239 & $-$20 08 30.19  &   44 & 560   \\
G240.32$+$0.07 &   4.72$\pm$0.47 &   8.3 $\times$ 10$^3$   &   07 44 52.040 &  $-$24 07 42.21 &  10 &  8473  \\
G359.97$-$0.46\tablefootmark{b}  &   {\it 4.0 }    &   {\it 5.7 $\times$ 10$^4$}     &  17 47 20.188 & $-$29 11 59.04  &   4 &   975   \\
{\it AFGL~5142 } &   2.14$\pm$0.05 &   5.0 $\times$ 10$^3$        &  05 30 48.018 & $+$33 47 54.54  &    23 &  469   \\
{\it IRAS 20126$+$4104} &   1.64$\pm$0.05 &   1.3 $\times$ 10$^4$    &   20 14 26.054 & $+$41 13 32.49   &  26 &  865   \\
\end{longtable}
\tablefoot{Column~1 reports the name of the POETS targets: the two sources in italic characters have been observed prior of the POETS survey; Cols.~2~and~3 give the trigonometric parallax distance from BeSSeL (available for all but four sources, for which the kinematic distance is reported) and the evaluated bolometric luminosity, respectively: kinematic distances and more uncertain luminosities are given in italic characters, and values which might be severe upper limits are enclosed within brackets; Cols.~4~and~5 give the RA and DEC coordinates, respectively, of the YSO, pinpointed by JVLA, high-angular resolution, radio continuum observations; Cols.~6~and~7 list the number of detected maser cloudlets and the maximum maser separation from the YSO, respectively.\\
\tablefoottext{a}{In a few cases where the continuum emission shows two nearby peaks, we choose the peak closer to most of the water masers. In three objects, G005.88$-$0.39, IRAS~20126$+$4104, and AFGL~5142, previously studied at subarcsecond resolution with millimeter and near-infrared interferometers, we have used the position of the YSO identified through continuum or molecular line emissions.}\\
\tablefoottext{b}{In this target, the water maser emission has faded away over the VLBI observing epochs, and nor trigonometric parallax distance neither reliable maser proper motions have been measured.}\\
\tablefoottext{c}{In this target, the water maser emission comes from two nearby YSOs, resolved through subarcsecond VLA observations, named VLA~3A and VLA~3B.}\\
\tablefoottext{d}{In this target, the water maser emission comes from two nearby YSOs, resolved through subarcsecond VLA observations, named VLA-1 and VLA-2.}
}
%\end{landscape}
%\end{table*}
%\end{sidewaystable*}
}

%--------------------------------------------------------------------
%\section{Observations}
\section{H$_2$O maser statistics}
\label{obs_res}

\subsection{Distance from the YSO}
\label{stat_dist}

In the POETS survey, on the basis of the spatial distribution of the water masers and their PMs, we could ascertain that 36 out of the 38 water masers observed with the JVLA are clearly associated with a radio continuum source. 
In Fig.~\ref{histo_rad}, we analyze the linear separation of the detected water masers from the position of the radio continuum peak.
In most of our targets, the continuum emission at both 13 and 22~GHz is dominated by a single (unresolved or slightly resolved) source, whose position is determined by fitting a 2D Gaussian profile. For each POETS target, we always select the radio continuum peak of the JVLA observation at the highest angular resolution, that is either 13~GHz or 22~GHz A-Array configuration. The histogram in the main plot of Fig.~\ref{histo_rad} considers all the water masers associated with the POETS targets, while the histogram in the inset only refers to masers up to a maximum separation of 3000~au from the continuum. Since the error on the linear distance is directly proportional to the distance at large values, the bins of the main and inset histograms are chosen 200~and~100~au, respectively, which is comparable with the average distance error of the selected masers. The histogram value is the fractional number of water masers in a bin. From these plots we infer that \  84\%  \ of the water masers are found within 1000~au from the radio continuum peak and \  45\% \ within \ 200~au. 

As discussed in Mos16, the peak of the radio continuum at 13 or 22~GHz is a good proxy for the YSO position, within the absolute position accuracy of $\approx$20~mas of JVLA observations. This belief is supported by several facts ascertained for the large majority of our targets: \ 1)~the peaks of the radio continuum at 6,~13,~and~22~GHz coincide in position within the errors and the determined radio spectral index, in the range of \ $-$0.1 -- 1.3, is consistent with thermal bremsstrahlung from a YSO wind; \ 2)~the water masers are distributed across the radio continuum at separations \ $\le$1000~au from the continuum peak; \ 3)~the water maser 3D velocities are best interpreted in terms of a DW emerging from the YSO (see Paper~III).
Therefore, on the basis of the distance distribution from the continuum shown in Fig.~\ref{histo_rad}, we are able to statistically assess for the first time that water masers can be efficiently used to pinpoint the position of a YSO with an uncertainty of a few 100~au. We discuss this result in the context of the relevant literature in Sect.~\ref{discu_dist}.  Table~\ref{wat_kin} lists the POETS targets, reporting distance, bolometric luminosity,  YSO position pinpointed by the radio continuum,  number of detected maser cloudlets, and the maximum maser separation from the YSO. 
%Furthermore, this result evidences that the separation of water masers from the central star is comparable (or lower) than the expected size of an accretion disk. 

\subsection{Isotropic luminosity}
\label{stat_lum}

Most of the water masers detected in the POETS survey have their peak intensity in the range of \ 0.05--50~\Jyb.
The isotropic luminosity, $L_{iso}$, of a maser cloudlet is calculated with the product \ $ L_{iso} = F_p \, D^2 $, where \ $F_p$ \ is the (Gaussian-fit) flux of the strongest emission channel, averaged over the different observing epochs, and \ $D$ \ is the source distance. 
The large majority, that is, 96\%, of the masers,  have \ $ L_{iso} \le 300$~Jy~kpc$^2$, although exceptionally bright masers with \ $ L_{iso}$ $\sim$$10^3$--$10^4$~Jy~kpc$^2$ \ have also been observed. The histograms in Fig.~\ref{histo_lum} show that the distribution of maser luminosity is strongly peaked towards low values: in considering masers with \ $ L_{iso} \le 300$~Jy~kpc$^2$, 86\% of them have  \ $ L_{iso} \le 50$~Jy~kpc$^2$; taking the masers with \ $ L_{iso} \le 30$~Jy~kpc$^2$, 56\% of them have  \ $ L_{iso} \le 5$~Jy~kpc$^2$. The implications of this luminosity distribution for more sensitive water maser observations of the future are considered in Sect.~\ref{discu_lum}.

\subsection{3D velocities}

\label{stat_vel}

\begin{figure}
\centering
\includegraphics[width=0.5\textwidth]{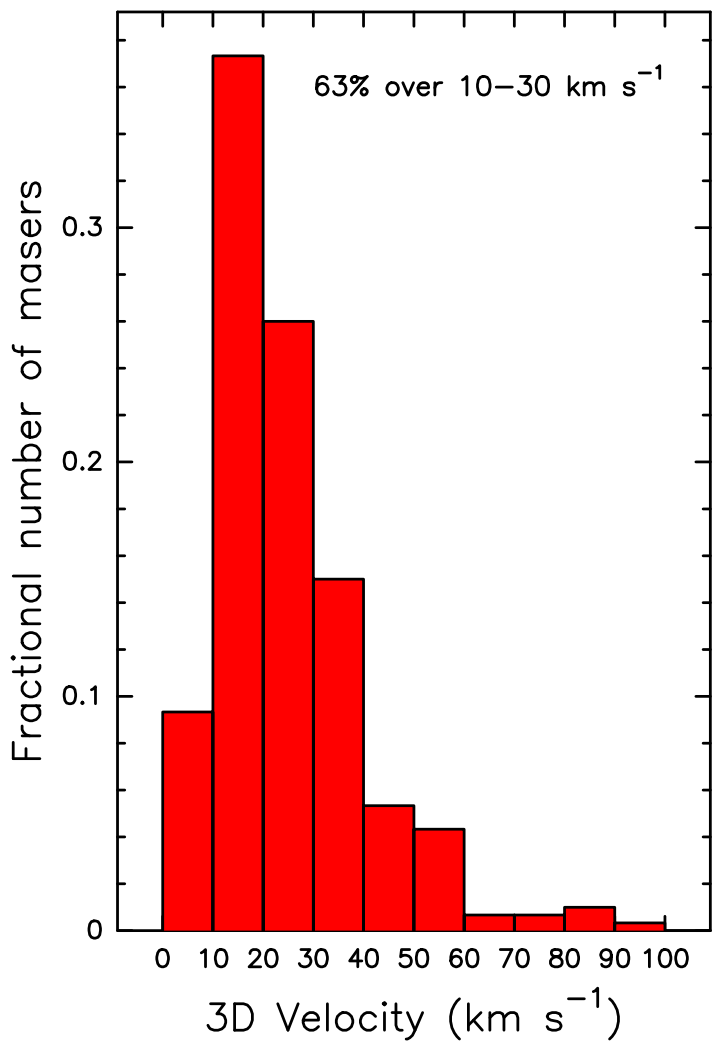}
\caption{Histogram of the maser 3D-velocity amplitude, cumulating all the POETS targets. The histogram bin is 10~\kms\ and the histogram values are normalized by the total number of masers with measured PMs. The average error on the 3D-velocity amplitude and corresponding standard deviation are \ 3.3~\kms \ and 2.5~\kms.}
\label{histo_v3d}
\end{figure}

\begin{figure}
\centering
\includegraphics[width=0.5\textwidth]{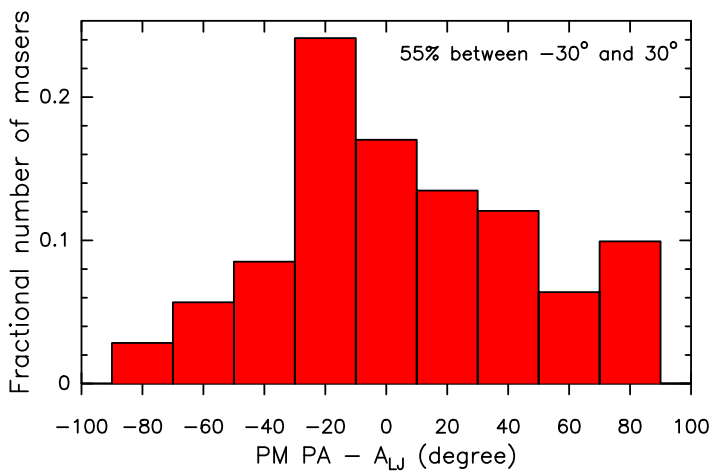}
\caption{Histogram of the difference between the maser PM~PA and the PA of the jet axis from the literature, A$_{\rm LJ}$, cumulating all the 14 POETS targets in which a jet has been observed through thermal tracers (see Table~A1 of Paper~III). The histogram bin is 20\degree\ and the histogram values are normalized by the total number of considered PMs. The target-average error on the PM~PA and corresponding standard deviation are\ 12\degree~and~4\degree. }
\label{histo_PA-LIT}
\end{figure}

%\begin{figure*}
%\centering
%\includegraphics[width=\textwidth]{/export/home/deva/mosca/BESSEL_FU/plot/PM_PA-POS_PA-npoi.eps}
%\caption{}
%\label{cont}
%\end{figure*}

\begin{figure}
\centering
\includegraphics[width=0.5\textwidth]{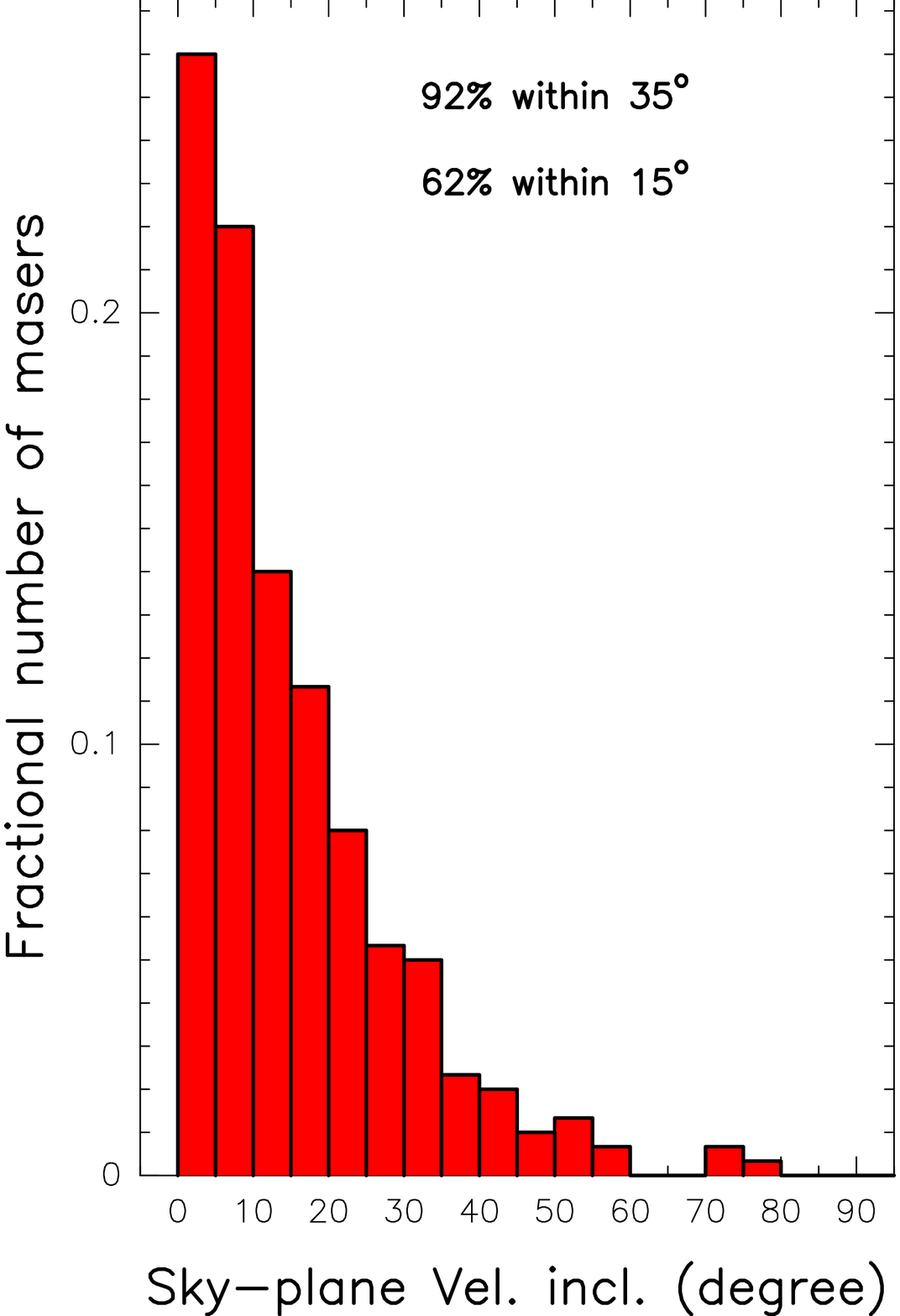}
\caption{Histogram of the inclination of the 3D water maser velocity with the plane of the sky, cumulating all the POETS targets. The histogram bin is 5\degree\  and the histogram values are normalized by the total number of masers with measured PMs. The average error on the maser velocity inclination and corresponding standard deviation are \ 7\degree~and~4\degree.}
\label{histo_incl}
\end{figure}

Combining the water maser PM and line of sight (LOS) velocity, we derive the maser 3D velocity. We have been searching the literature for the systemic \Vlsr\ of our targets from observations of high-density (n$_{\rm{H}_2} \ge 10^6$~cm$^{-3}$) thermal tracers and, in most cases, the found values are accurate within a few \kms. Since PMs are determined with comparable accuracies of a few \kms \ (see Tables~6--16 in Mos16, and Tables~B1--B27 in Paper~III), we estimate an average error on the 3D-velocity amplitude of \ 3.3~\kms. The histogram of the maser 3D-velocity amplitudes is presented in Fig.~\ref{histo_v3d}, from which it is clear that a large fraction, 63\%, of the water maser speeds falls in the range \ 10--30~\kms. 

It is also interesting to study the angular distribution of the maser 3D velocities, considering both their projection into the plane of the sky and their inclination with respect to it.
In Paper~III (see Figs.~C1--C3), we produced histograms of the position angle (PA) of the water maser PMs for all the targets with at least two measured PMs and studied the degree of collimation of the maser velocities in the plane of the sky. In searching the literature for high-angular ($\la$1\arcsec) resolution observations in thermal continuum and line tracers of our targets, for several cases we were able to ascertain the PA of the (sky-projected) axis of the YSO jet, referred to as A$_{\rm LJ}$ (see Paper~III, Table~A1). Figure~\ref{histo_PA-LIT} presents the histogram of the difference between the maser \ PM~PA \ and \ A$_{\rm LJ}$, cumulating all the 14 POETS targets for which the direction of the protostellar jet is known from the literature (see Paper~III, Table~A1). The average error on the PM~PA is about half of the histogram bin. We note that \ 55\% \ of the PMs of the masers is directed on the sky within an angle of \ 30\degree\ from the jet axis. 

In Fig.~\ref{histo_incl} we have plotted the histogram of the inclination of the 3D velocities with the plane of the sky for all the water masers with measured PMs. The histogram bin is comparable with the average error of the inclination. It is clear that most of the observed water masers move close to the plane of the sky, 92\% \ and \ 62\% \ within an angle of \ 35\degree\ and 15\degree, respectively.

\section{Discussion}
\label{discu}

\subsection{H$_2$O masers marking the YSOs}
\label{discu_dist}

The result presented in Sect.~\ref{stat_dist} is derived from absolute positions (accurate to a few~mas) of \  994 \ individual maser cloudlets associated with 36 distinct YSOs  (see Table~\ref{wat_kin}), whose position on the sky and distance are determined with an accuracy of a few 10~mas and \ 5--10\% (via maser trigonometric parallax measurements), respectively. The mas-level accuracy in our VLBI H$_2$O maser positions together with the employment of a reliable YSO tracer (i.e., sensitive, rms$\sim$10~$\mu$\Jyb, JVLA A-Array observations) distinguishes ours from previous surveys of interstellar water masers, such as those by \citet{Beu02a}, \citet{Wal14}, \citet{Tit16}, and \citet{Kim19}. These former surveys were conducted with connected element interferometers (Very Large Array, VLA; Australian Telescope Compact Array, ATCA; Korean VLBI Network, KVN) achieving an angular resolution of \ $\ge$~0\farcs1, which is insufficient for determining the spatial and velocity distribution of individual maser cloudlets. 

%Besides, these large surveys have targeted various high-mass star formation indicators, as cm~and~mm~continuum sources, 6.7~GHz methanol masers, and ultracompact (UC) \HII~regions, but they are not complemented with specific observations to search for a YSO near the water masers themselves. A significant fraction of the water masers is always found to be offset by a few arcsec from the targets, 
In these previous surveys, a significant fraction of the water masers is always found to be offset by a few arcsec from the complementary star formation indicators (such as mm-continuum sources and 6.7~GHz CH$_3$OH masers), calling into question the connection between the water masers and a specific phase of the YSO evolution. 
%Given the very clustered environment of high-mass star formation as revealed, for instance, by recent ALMA subarcsecond observations \citep{Beu19}, we consider it very plausible that the water masers not associated with the targets of these surveys are powered by nearby, undetected cluster components. 
This apparent discrepancy can be readily explained by limitations of previous surveys, including poor resolution and sensitivity (typically an order of magnitude worse  
than POETS for the radio continuum).
At 22~GHz, the radio continuum sensitivity of the POETS survey ($\sim$30~$\mu$Jy over a beam size of \ 0\farcs1) allows us to detect the flux from an unresolved, optically thin, \HII~region excited by a zero-age main-sequence (ZAMS) star with spectral type as late as B5 at a distance of \ 1~kpc, and \ B3--B4 \ at \ 3--4~kpc. 
Therefore, sensitive JVLA A-Array continuum observations towards the water masers detected in these previous surveys would provide a critical test of our results by searching for previously undetected cluster components emitting (free-free) radio continuum. Indeed, our results clearly indicate a strong connection between the water masers and an active
%, indicating instead a strong connection between the  water masers and the most active 
phase of mass ejection (and accretion) of the YSO.

\subsection{The H$_2$O maser brightness function}
\label{discu_lum}

We do not find any clear correlation between the intensity of the water maser and its distance from the YSO or its 3D-velocity amplitude. In general, water masers cluster on scales of \ $\sim$10~au and it is quite common to find mixed groups of intense and weak masers at similar distance from the YSO. This qualitatively agrees with the shock models for the origin of the H$_2$O masers, which predict that the peak flux of the masers depends mostly on the LOS velocity coherent, amplification path \citep[][Eq.~14]{Hol13} in the post-shock gas layers. Then, a large range of maser amplification paths (and intensities) across small regions is naturally expected since:\ 1)~both the density and velocity behind shocks vary on very small scales, namely, $\le$1~au \citep[][see Fig.~4]{Hol13}; 2)~the shock velocities can be inclined at different angles with respect to the LOS; \ 3)~depending on the very local conditions of the ambient medium, shocks can even become fragmented and dissipate. In this way, the scatter in maser intensities owing to the specific shock geometry can conceal a weaker dependence of the maser intensity on the position from the YSO and the shock speed.

The histograms of Fig.~\ref{histo_lum} indicate that the channel-map detection threshold of  \ $\sim$100~m\Jyb \ of present spectroscopic VLBI measurements is insufficient to adequately image the water maser emission associated with YSOs. In fact, the steep increase of the fractional number of detected masers with decreasing luminosity strongly suggests that there is a multitude of masers with an intensity \ $\le$100~m\Jyb \ remaining undetected. In Fig.~\ref{histo_lum}, we report the power-law fits of the histogram values both for the main plot and the inset. We have to consider that the first bins of these histograms are certainly underestimated owing to our limited sensitivity. It should also be noted that the fitted power-law expressions reported in Fig.~\ref{histo_lum} are not normalized. Requiring that the integral between  \ $0 \le  L_{iso} \le 30$~Jy~kpc$^2$ \ be equal to \ 1, we determine the
corrected brightness distribution for weak water masers as: 
\begin{equation}
\label{eq_wmbd}
  F_{\rm M} = 0.11 \; (L_{iso})^{-0.70} 
.\end{equation}

Equation~\ref{eq_wmbd} implies that about 50\% of the whole water maser population have a luminosity that is less than the typical detection threshold of a few~Jy~kpc$^2$ and it is still undetected. 
 The next-generation VLA (ngVLA) will be the appropriate facility to study the spatial and velocity distribution of this large family of weaker masers \citep{Hun18}, which can be used to sample the 3D velocity field around the forming YSO with much better detail. If water masers at a distance of \ $\la$1000~au from the YSOs trace magneto-centrifugal DWs (see Paper~III), multi-epoch ngVLA, polarized 22~GHz maser observations 
 can serve as a unique tool to investigate both the velocity and magnetic fields of DWs. Moreover, the ngVLA will permit a simultaneous study of the water masers and the (usually faint) radio continuum of the YSO, improving the precision of the maser positions (and velocities) in the YSO reference frame.  

\subsection{The origin of H$_2$O masers in shocks}
\label{discu_vel}

\begin{figure}
%\centering
\includegraphics[width=0.5\textwidth]{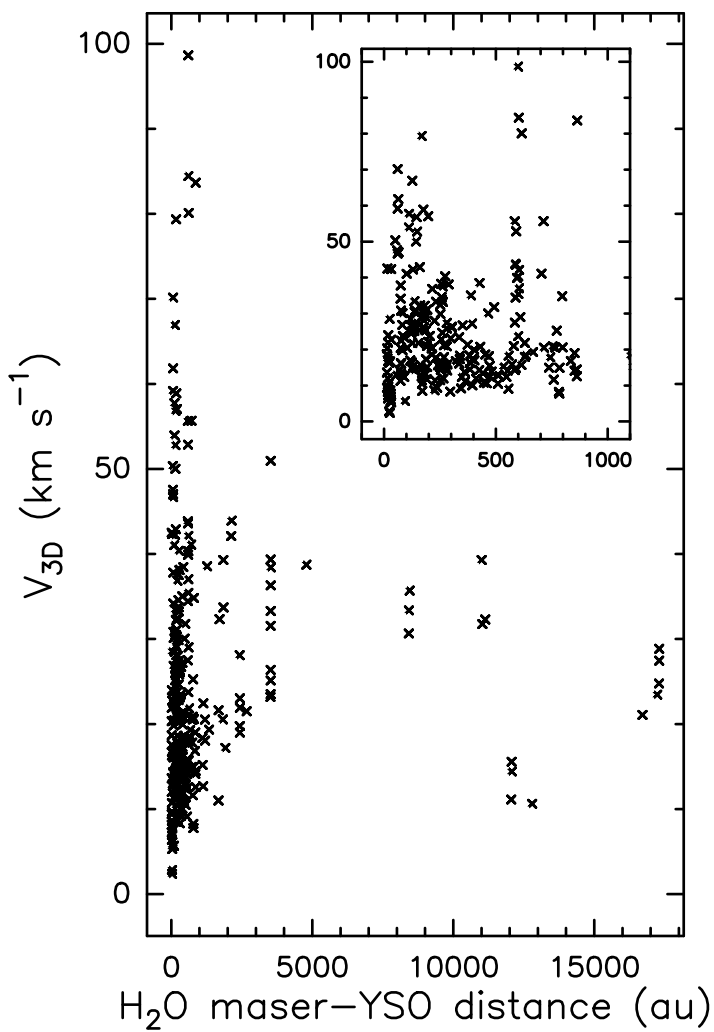}
\caption{Plot of the amplitude of the 3D velocities of the H$_2$O masers vs. the distance from the YSO, cumulating all the POETS targets. The inset in the upper~right corner shows an enlargement of the main plot for the distance range \ 0--1100~au. }
%\caption{Plot of the amplitude of the 3D ({\em upper-panel}) and LOS ({\em lower-panel}) velocities of the H$_2$O masers vs. the distance from the YSO, cumulating all the POETS targets. In each panel, the inset in the upper~right corner shows an enlargement of the main plot for the distance range \ 0--1100~au. }
\label{V_dist}
\end{figure}

Concerning the statistics on the properties of the H$_2$O maser 3D velocities, the first notable result is the global distribution of the PM directions with respect to the  orientation of the YSO jet, presented in Fig.~\ref{histo_PA-LIT}. The fact that more than half of the observed PMs is oriented on the sky within an angle of \ 30\degree\ from the jet axis implies that, statistically, the observation of only a few maser PMs associated with a given YSO can already suffice to constrain the direction of the protostellar jet. In the literature, there have been several studies published on specific sources assessing that water masers can be reliable tracers of the YSO outflows (see Sect.~\ref{intro} for references), but our H$_2$O maser survey first provides a statistic of the water maser PM directions based on a relatively large sample of targets. Fig.~\ref{histo_PA-LIT} can also be read that about half of the water masers move along directions forming large angles with the jet. The non-collimated motion, near the YSO, of half of the water masers has been interpreted in Paper~III in terms of a wide-angle velocity pattern associated
% with the DW powering the jet.
with the outer, less collimated portion of the DW.

The second interesting result is the finding that almost all the measured maser 3D velocities form small angles ($\le$ 35\degree) with the plane of the sky, as indicated by the histogram in Fig.~\ref{histo_incl}. That agrees very well with the shock models for the water masers 
\citep{Eli92,Kau96,Hol13}, where a long velocity-coherent, amplification path along the LOS is predicted 
%for planar shocks being accelerated and moving close to the plane of the sky. In this way, the velocity change across the maser due to ordered motions is small compared with the thermal or microturbulent line width. 
for planar shocks with propagation direction close to the plane of the sky. In this case, there are no large variations of the velocity along the line of sight, which permits the formation of intense masers.
Therefore, we can explain the histograms in Fig.~\ref{histo_incl} as an observational bias, given that masers moving at small (large) angles from the plane of the sky have a much higher (lower) detection probability. 

In the following, we discuss in more detail the properties of the shocks excited in protostellar winds that can serve as the birthplaces of the H$_2$O masers. Let us indicate with \ $V_{\rm w}$ \ and \ $\rho_{\rm w}$ \ the speed and the density, respectively, of the wind, and with \ $V_{\rm u}$ \ 
and \ $\rho_{\rm u}$ \ the speed and the density, respectively, of the pre-shock material on which the wind impinges, that is \ $V_{\rm w} \ge V_{\rm u}$. Both for radiative and adiabatic shocks, the shock velocity, $ V_{\rm s} $, approximated by the water maser velocity, can be written as \citep[][see Eqs.~3~and~4]{Mas93}: 

\begin{equation} 
\label{eq_vs}
 V_{\rm s} \approx V_{\rm u} + \frac{V_{\rm w} - V_{\rm u}}{1+\sqrt{\rho_{\rm u} / \rho_{\rm w} }} 
 .\end{equation}
   
According to the shock models, H$_2$O masers originate in post-shock gas layers with (H$_2$ number) density of \ n$_{\rm{H}_2} \sim 10^8$--$10^9$~cm$^{-3}$ \citep{Hol13}. These high post-shock densities can be obtained either behind weak C~shocks propagating through very dense regions, 
$ \rho_{\rm u}  \sim $ $10^8$--$10^9$~cm$^{-3}$, $5 \le (V_{\rm s} - V_{\rm u}) \le 40$~\kms, or behind strong J~shocks crossing less dense gas, $ \rho_{\rm u}  \sim $ $10^6$--$10^7$~cm$^{-3}$, $ (V_{\rm s} - V_{\rm u}) \ge 50$~\kms.

Gas densities as high as \ $10^8$--$10^9$~cm$^{-3}$ \ are typical of the few disks discovered around high-mass YSOs 
\citep{Beu13,Joh15,Ile16,Mos19,San19a}. DWs are expected to be launched from the surface of such dense disks \citep{Pud05,Koe18}. On the basis of Eq.~\ref{eq_vs}, weak, internal C~shocks in the slower and less collimated regions of DWs, with \ $\rho_{\rm w} \approx \rho_{\rm u}  \sim $ $10^8$--$10^9$~cm$^{-3} $, and \ 10~\kms $ \le V_{\rm u} \la V_{\rm w} \le 40$~\kms, would propagate at speeds of \ 10--30~\kms, in good agreement with the peak of the velocity distribution of the water masers shown in Fig.~\ref{histo_v3d}. 
%Therefore, the analysis of the direction and amplitude of the velocities suggests that a large fraction of the water masers at smaller separation from the YSOs originate in C~shocks internal to DWs.
Therefore, based on the analysis of the direction and amplitude of the maser velocities, we suggest that about 50\% \ of the water masers, excited within 1000~au from the YSOs, originate in C~shocks internal to DWs.

The highest maser speeds, $\ge$30~\kms, of the distribution of Fig.~\ref{histo_v3d} could instead be generated in strong J~shocks, which are naturally expected to be associated with fast protostellar jets. The radio continuum emission from thermal jets is thought to be produced in internal shocks and provides a reliable way to estimate the jet properties. At \ 100--1000~au from the YSOs, typical jet speeds and densities are \ $ V_{\rm w} \approx$ 100--1000~\kms \ and \ $ \rho_{\rm w}$~$\sim$ $10^6$~cm$^{-3}$, respectively \citep{Ang18,Ros19}. If such fast jets impact denser, $\rho_{\rm u} \approx$ $10^7$--$10^8$~cm$^{-3}$, stationary, $V_{\rm u} \approx$ 0~\kms, ambient material, Eq.~\ref{eq_vs} predicts velocities for the shocks a factor of \ 3-10 \ smaller than that of the jet, in agreement with our measurements for the faster subset of water masers. Thus, we propose that the water masers showing higher velocities, that is, $\ge$30~\kms, or larger separation, at several 1000~au, from the YSO originate in J~shocks owing to the interaction of the protostellar jet with the surrounding material.

Figure~\ref{V_dist} presents the distribution of the 3D velocity amplitude of the water masers versus the distance from the YSO. In particular, the plot in the inset shows that the large majority of the masers within 1000~au have speeds \ $\le$30~\kms, and, following our interpretation, it should originate in internal C~shocks of the YSOs DWs. Considering, instead, the near, $\le$1000~au, fastest, $\ge$30~\kms, masers, we note that a significant fraction of them are found within \ 200~au from the YSO. The plot of Fig.~\ref{V_dist} cumulates all the POETS targets, covering a large range in bolometric luminosities, 10$^2$--10$^5$~\ls, and YSO masses, 5--30~\ms. A corresponding large variation in the sizes of the kinematic structures (i.e., disks, DWs and jets) surrounding the YSOs is also expected. Then, the plot region \ $\le$~200~au~and~$\ge$~30~\kms \ could correspond to water masers associated with the roots of the jets in lower-mass YSOs.  Inspecting the main plot of Fig.~\ref{V_dist}, it is also remarkable that the minimum water maser velocity increases regularly with the distance from the YSO up to, at least, \ 4000--5000~au. Following our discussion, at separations of \ 1000--10000~au, water masers should mainly arise in J~shocks excited in the jet-ambient interaction, and, on the basis of Eq.~\ref{eq_vs}, a plausible interpretation is that the shock (and maser) velocities increase because the jet is continuously accelerated at this range of distances \citep{Pud07,Ram19}. 

%In Sects.~\ref{stat_dist}~and~\ref{discu_dist}, we have presented strong evidence that a large fraction of the water masers arises within a few 100~au from the YSOs, and in Sect.~\ref{stat_vel}, we have shown that the probability of maser detection enhances if the masers move closer and closer to the plane of the sky, or, equivalently, if the ratio of the LOS and sky-projected velocity components is smaller and smaller. Based on these previous results, the plot in the lower~panel of Fig.~\ref{V_dist} suggests a criterion to judge the YSO-maser distance based only on the LOS maser velocity. In fact, owing to the larger statistics, at smaller separations from the YSO the spread in LOS velocity is much higher: from \ $\le$~5~\kms\ at distances \ $\ge$5000~au, it raises to \ $\ge$~30~\kms\ at distances of a few 100~au. From our survey, we derive that \ 10\% \ of the detected water masers have LOS velocity higher than \ 10~\kms. Combining this figure with the derived probability of finding masers at separation \ $\ge$1000~au, 17\% \ (see Sect.~\ref{stat_dist}, we can state that the probability of detecting distant ($\ge$1000~au), highly red-~or~blue-shifted ($\ge$10~\kms) water masers is very low, 1.7\%. Thus, if, in general, water masers are a good proxy for the position of the YSO (see Sect.~\ref{discu_dist}, those at high red-~or~blue-shifts are even more reliable. 

\section{Conclusions}
\label{conclu}

%This paper presents statistical information on the properties of a sample (36) of H$_2$O masers extracted from the list of targets of the BeSSeL VLBA-key project. 
This paper presents a statistical analysis of the kinematic properties at mas resolution of a large sample of \  994 \ H$_2$O masers, associated with 36 distinct YSOs.
To our knowledge, this is the first VLBI survey of the 22~GHz H$_2$O masers, including a sizable number of sources and based on maser positions and 3D velocities determined with an accuracy of a few mas and a few \kms, respectively. The main results can be summarized as follows:
\begin{enumerate}
\item The 22~GHz water masers are a reliable proxy for the position of the YSOs: 84\% of them are found within 1000~au from the YSO, and \  45\%  \ within \ 200~au. 

\item The distribution of maser luminosity is strongly peaked towards low values, indicating that about half of the maser population, with  intensities \ $\le$100~m\Jyb, remains undetected.

\item The PA of the water maser PMs provides a statistically-significant estimate for the PA of the YSO jet: 55\% \ of the maser PMs are directed on the sky within an angle of \ 30\degree\ from the jet axis.

\item The 3D maser velocities have small ($\le$35\degree) inclinations with the plane of the sky and amplitudes mainly in the range of 10--30~\kms.

\end{enumerate}  

Regarding point~2, future sensitive interferometers (e.g., ngVLA) will serve as the appropriate instruments for studying the weaker, albeit conspicuous, water maser population predicted in this work.
Concerning points~3~and~4, we stress that for the first time we are providing a statistic for the amplitude and directions of 3D velocities for the 22~GHz water masers. We find that water masers moving almost parallel to the plane of the sky are preferentially detected. This result fully supports the shock models for the origin of the water masers, where a long velocity-coherent, amplification path along the LOS is predicted for planar shocks with propagation direction close to the plane of the sky. In Paper~III, which focused on the maser kinematics, we proposed that the water masers near the YSOs trace the different regions of a DW, from the fast and collimated jet portion to the slower, wide-angle outer layers. Following this interpretation, we can now make a prediction on the nature of the maser shocks. While most of the water masers with lower speed, that is, $\le$30~\kms, should be excited in weak, internal C~shocks occurring in the slower regions of the DWs, most of the masers at higher velocity, that is, $\ge$30~\kms, should originate in strong J~shocks wherever the fast jets hit the ambient medium.

%\begin{acknowledgements}

%\end{acknowledgements}

   \bibliographystyle{aa} % style aa.bst
   \bibliography{biblio} % your references Yourfile.bib

\begin{thebibliography}{39}
\expandafter\ifx\csname natexlab\endcsname\relax\def\natexlab#1{#1}\fi

\bibitem[{{Anglada} {et~al.}(1996){Anglada}, {Estalella}, {Pastor},
  {Rodriguez}, \& {Haschick}}]{Ang96b}
{Anglada}, G., {Estalella}, R., {Pastor}, J., {Rodriguez}, L.~F., \&
  {Haschick}, A.~D. 1996, \apj, 463, 205

\bibitem[{{Anglada} {et~al.}(2018){Anglada}, {Rodr{\'\i}guez}, \&
  {Carrasco-Gonz{\'a}lez}}]{Ang18}
{Anglada}, G., {Rodr{\'\i}guez}, L.~F., \& {Carrasco-Gonz{\'a}lez}, C. 2018,
  \aapr, 26, 3

\bibitem[{{Beuther} {et~al.}(2013){Beuther}, {Linz}, \& {Henning}}]{Beu13}
{Beuther}, H., {Linz}, H., \& {Henning}, T. 2013, \aap, 558, A81

\bibitem[{{Beuther} {et~al.}(2002){Beuther}, {Walsh}, {Schilke}, {Sridharan},
  {Menten}, \& {Wyrowski}}]{Beu02a}
{Beuther}, H., {Walsh}, A., {Schilke}, P., {et~al.} 2002, \aap, 390, 289

\bibitem[{{Burns} {et~al.}(2016){Burns}, {Handa}, {Nagayama}, {Sunada}, \&
  {Omodaka}}]{Bur16}
{Burns}, R.~A., {Handa}, T., {Nagayama}, T., {Sunada}, K., \& {Omodaka}, T.
  2016, \mnras, 460, 283

\bibitem[{{Cheung} {et~al.}(1969){Cheung}, {Rank}, {Townes}, {Thornton}, \&
  {Welch}}]{Che69}
{Cheung}, A.~C., {Rank}, D.~M., {Townes}, C.~H., {Thornton}, D.~D., \& {Welch},
  W.~J. 1969, \nat, 221, 626

\bibitem[{{Elitzur} {et~al.}(1989){Elitzur}, {Hollenbach}, \& {McKee}}]{Eli89}
{Elitzur}, M., {Hollenbach}, D.~J., \& {McKee}, C.~F. 1989, \apj, 346, 983

\bibitem[{{Elitzur} {et~al.}(1992){Elitzur}, {Hollenbach}, \& {McKee}}]{Eli92}
{Elitzur}, M., {Hollenbach}, D.~J., \& {McKee}, C.~F. 1992, \apj, 394, 221

\bibitem[{{Felli} {et~al.}(2007){Felli}, {Brand}, {Cesaroni}, {Codella},
  {Comoretto}, {Di Franco}, {Massi}, {Moscadelli}, {Nesti}, {Olmi}, {Palagi},
  {Panella}, \& {Valdettaro}}]{Fel07}
{Felli}, M., {Brand}, J., {Cesaroni}, R., {et~al.} 2007, \aap, 476, 373

\bibitem[{{Furuya} {et~al.}(2001){Furuya}, {Kitamura}, {Wootten}, {Claussen},
  \& {Kawabe}}]{Fur01}
{Furuya}, R.~S., {Kitamura}, Y., {Wootten}, H.~A., {Claussen}, M.~J., \&
  {Kawabe}, R. 2001, \apjl, 559, L143

\bibitem[{{Goddi} \& {Moscadelli}(2006)}]{God06a}
{Goddi}, C. \& {Moscadelli}, L. 2006, \aap, 447, 577

\bibitem[{{Goddi} {et~al.}(2006){Goddi}, {Moscadelli}, {Torrelles}, {Uscanga},
  \& {Cesaroni}}]{God06b}
{Goddi}, C., {Moscadelli}, L., {Torrelles}, J.~M., {Uscanga}, L., \&
  {Cesaroni}, R. 2006, \aap, 447, L9

\bibitem[{{Hollenbach} {et~al.}(2013){Hollenbach}, {Elitzur}, \&
  {McKee}}]{Hol13}
{Hollenbach}, D., {Elitzur}, M., \& {McKee}, C.~F. 2013, \apj, 773, 70

\bibitem[{{Hunter} {et~al.}(2018){Hunter}, {Brogan}, {Bartkiewicz}, {Chibueze},
  {Cyganowski}, {Hirota}, {MacLeod}, {Sanna}, \& {Torrelles}}]{Hun18}
{Hunter}, T.~R., {Brogan}, C.~L., {Bartkiewicz}, A., {et~al.} 2018,
  Astronomical Society of the Pacific Conference Series, Vol. 517,
  {Understanding Massive Star Formation through Maser Imaging}, ed.
  E.~{Murphy}, 321

\bibitem[{{Ilee} {et~al.}(2016){Ilee}, {Cyganowski}, {Nazari}, {Hunter},
  {Brogan}, {Forgan}, \& {Zhang}}]{Ile16}
{Ilee}, J.~D., {Cyganowski}, C.~J., {Nazari}, P., {et~al.} 2016, \mnras, 462,
  4386

\bibitem[{{Johnston} {et~al.}(2015){Johnston}, {Robitaille}, {Beuther}, {Linz},
  {Boley}, {Kuiper}, {Keto}, {Hoare}, \& {van Boekel}}]{Joh15}
{Johnston}, K.~G., {Robitaille}, T.~P., {Beuther}, H., {et~al.} 2015, \apjl,
  813, L19

\bibitem[{{Kaufman} \& {Neufeld}(1996)}]{Kau96}
{Kaufman}, M.~J. \& {Neufeld}, D.~A. 1996, \apj, 456, 250

\bibitem[{{Kim} {et~al.}(2019){Kim}, {Kim}, \& {Kim}}]{Kim19}
{Kim}, W.-J., {Kim}, K.-T., \& {Kim}, K.-T. 2019, \apjs, 244, 2

\bibitem[{{K{\"o}lligan} \& {Kuiper}(2018)}]{Koe18}
{K{\"o}lligan}, A. \& {Kuiper}, R. 2018, \aap, 620, A182

\bibitem[{{Masson} \& {Chernin}(1993)}]{Mas93}
{Masson}, C.~R. \& {Chernin}, L.~M. 1993, \apj, 414, 230

\bibitem[{{Moscadelli} {et~al.}(2005){Moscadelli}, {Cesaroni}, \&
  {Rioja}}]{Mos05}
{Moscadelli}, L., {Cesaroni}, R., \& {Rioja}, M.~J. 2005, \aap, 438, 889

\bibitem[{{Moscadelli} {et~al.}(2016){Moscadelli}, {S{\'a}nchez-Monge},
  {Goddi}, {Li}, {Sanna}, {Cesaroni}, {Pestalozzi}, {Molinari}, \&
  {Reid}}]{Mos16}
{Moscadelli}, L., {S{\'a}nchez-Monge}, {\'A}., {Goddi}, C., {et~al.} 2016,
  \aap, 585, A71

\bibitem[{{Moscadelli} {et~al.}(2019{\natexlab{a}}){Moscadelli}, {Sanna},
  {Cesaroni}, {Rivilla}, {Goddi}, \& {Rygl}}]{Mos19}
{Moscadelli}, L., {Sanna}, A., {Cesaroni}, R., {et~al.} 2019{\natexlab{a}},
  \aap, 622, A206

\bibitem[{{Moscadelli} {et~al.}(2019{\natexlab{b}}){Moscadelli}, {Sanna},
  {Goddi}, {Krishnan}, {Massi}, \& {Bacciotti}}]{Mos19b}
{Moscadelli}, L., {Sanna}, A., {Goddi}, C., {et~al.} 2019{\natexlab{b}}, \aap,
  631, A74

\bibitem[{{Pudritz} \& {Banerjee}(2005)}]{Pud05}
{Pudritz}, R.~E. \& {Banerjee}, R. 2005, in IAU Symposium, Vol. 227, Massive
  Star Birth: A Crossroads of Astrophysics, ed. R.~{Cesaroni}, M.~{Felli},
  E.~{Churchwell}, \& M.~{Walmsley}, 163--173

\bibitem[{{Pudritz} {et~al.}(2007){Pudritz}, {Ouyed}, {Fendt}, \&
  {Brandenburg}}]{Pud07}
{Pudritz}, R.~E., {Ouyed}, R., {Fendt}, C., \& {Brandenburg}, A. 2007, in
  Protostars and Planets V, ed. B.~{Reipurth}, D.~{Jewitt}, \& K.~{Keil}, 277

\bibitem[{{Ramsey} \& {Clarke}(2019)}]{Ram19}
{Ramsey}, J.~P. \& {Clarke}, D.~A. 2019, \mnras, 484, 2364

\bibitem[{{Reid} {et~al.}(2019){Reid}, {Menten}, {Brunthaler}, {Zheng}, {Dame},
  {Xu}, {Li}, {Sakai}, {Wu}, {Immer}, {Zhang}, {Sanna}, {Moscadelli}, {Rygl},
  {Bartkiewicz}, {Hu}, {Quiroga-Nunez}, \& {van Langevelde}}]{Rei19}
{Reid}, M.~J., {Menten}, K.~M., {Brunthaler}, A., {et~al.} 2019, arXiv
  e-prints, arXiv:1910.03357

\bibitem[{{Reid} {et~al.}(2014){Reid}, {Menten}, {Brunthaler}, {Zheng}, {Dame},
  {Xu}, {Wu}, {Zhang}, {Sanna}, {Sato}, {Hachisuka}, {Choi}, {Immer},
  {Moscadelli}, {Rygl}, \& {Bartkiewicz}}]{Rei14}
{Reid}, M.~J., {Menten}, K.~M., {Brunthaler}, A., {et~al.} 2014, \apj, 783, 130

\bibitem[{{Rosero} {et~al.}(2019){Rosero}, {Hofner}, {Kurtz}, {Cesaroni},
  {Carrasco-Gonz{\'a}lez}, {Araya}, {Rodr{\'\i}guez}, {Menten}, {Wyrowski},
  {Loinard}, {Ellingsen}, \& {Molinari}}]{Ros19}
{Rosero}, V., {Hofner}, P., {Kurtz}, S., {et~al.} 2019, \apj, 880, 99

\bibitem[{{Sanna} {et~al.}(2019{\natexlab{a}}){Sanna}, {K{\"o}lligan},
  {Moscadelli}, {Kuiper}, {Cesaroni}, {Pillai}, {Menten}, {Zhang}, {Caratti o
  Garatti}, {Goddi}, {Leurini}, \& {Carrasco-Gonz{\'a}lez}}]{San19a}
{Sanna}, A., {K{\"o}lligan}, A., {Moscadelli}, L., {et~al.} 2019{\natexlab{a}},
  \aap, 623, A77

\bibitem[{{Sanna} {et~al.}(2019{\natexlab{b}}){Sanna}, {Moscadelli}, {Goddi},
  {Beltr{\'a}n}, {Brogan}, {Caratti o Garatti}, {Carrasco-Gonz{\'a}lez},
  {Hunter}, {Massi}, \& {Padovani}}]{San19b}
{Sanna}, A., {Moscadelli}, L., {Goddi}, C., {et~al.} 2019{\natexlab{b}}, \aap,
  623, L3

\bibitem[{{Sanna} {et~al.}(2018){Sanna}, {Moscadelli}, {Goddi}, {Krishnan}, \&
  {Massi}}]{San18}
{Sanna}, A., {Moscadelli}, L., {Goddi}, C., {Krishnan}, V., \& {Massi}, F.
  2018, \aap, 619, A107

\bibitem[{{Sanna} {et~al.}(2012){Sanna}, {Reid}, {Carrasco-Gonz{\'a}lez},
  {Menten}, {Brunthaler}, {Moscadelli}, \& {Rygl}}]{San12}
{Sanna}, A., {Reid}, M.~J., {Carrasco-Gonz{\'a}lez}, C., {et~al.} 2012, \apj,
  745, 191

\bibitem[{{Titmarsh} {et~al.}(2016){Titmarsh}, {Ellingsen}, {Breen}, {Caswell},
  \& {Voronkov}}]{Tit16}
{Titmarsh}, A.~M., {Ellingsen}, S.~P., {Breen}, S.~L., {Caswell}, J.~L., \&
  {Voronkov}, M.~A. 2016, \mnras, 459, 157

\bibitem[{{Torrelles} {et~al.}(2003){Torrelles}, {Patel}, {Anglada}, {G{\'
  o}mez}, {Ho}, {Lara}, {Alberdi}, {Cant{\' o}}, {Curiel}, {Garay}, \&
  {Rodr{\'{\i}}guez}}]{Tor03}
{Torrelles}, J.~M., {Patel}, N.~A., {Anglada}, G., {et~al.} 2003, \apjl, 598,
  L115

\bibitem[{{Urquhart} {et~al.}(2011){Urquhart}, {Morgan}, {Figura}, {Moore},
  {Lumsden}, {Hoare}, {Oudmaijer}, {Mottram}, {Davies}, \& {Dunham}}]{Urq11}
{Urquhart}, J.~S., {Morgan}, L.~K., {Figura}, C.~C., {et~al.} 2011, \mnras,
  418, 1689

\bibitem[{{Walsh} {et~al.}(2011){Walsh}, {Breen}, {Britton}, {Brooks},
  {Burton}, {Cunningham}, {Green}, {Harvey-Smith}, {Hindson}, {Hoare},
  {Indermuehle}, {Jones}, {Lo}, {Longmore}, {Lowe}, {Phillips}, {Purcell},
  {Thompson}, {Urquhart}, {Voronkov}, {White}, \& {Whiting}}]{Wal11}
{Walsh}, A.~J., {Breen}, S.~L., {Britton}, T., {et~al.} 2011, \mnras, 416, 1764

\bibitem[{{Walsh} {et~al.}(2014){Walsh}, {Purcell}, {Longmore}, {Breen},
  {Green}, {Harvey-Smith}, {Jordan}, \& {Macpherson}}]{Wal14}
{Walsh}, A.~J., {Purcell}, C.~R., {Longmore}, S.~N., {et~al.} 2014, \mnras,
  442, 2240

\end{thebibliography}

%--------------------------------------------------------------------

\end{document}